\documentclass{article}
\usepackage{spconf,amsmath,graphicx}

\usepackage{siunitx}
\usepackage{xcolor}

\usepackage{pifont}
\newcommand{\cmark}{\ding{51}}%
\newcommand{\xmark}{\ding{55}\,}%

\usepackage[acronym, toc, nonumberlist]{glossaries}

\usepackage{amsfonts}
\usepackage{booktabs}
\usepackage{cleveref}
\usepackage{tikz}
\usetikzlibrary{positioning, shapes, calc, arrows}  
\usepackage{relsize}
\usepackage{multirow}
\usepackage{pgfplots}
\pgfplotsset{compat=newest}
\usepackage{placeins}
\usepackage{url}
\usepackage{balance}
\usepackage{stackengine}
\usepackage{paralist}
\usepackage{scrextend}

\usepackage[skip=4pt]{caption}

\newacronym{AdaIN}{AdaIN}{adaptive instance normalization}

\newacronym{BLSTM}{BLSTM}{bidirectional long short-term memory}
\newacronym{BN}{BN}{batch normalization}

\newacronym{CPC}{CPC}{contrastive predictive coding}
\newacronym{CNN}{CNN}{convolutional neural network}

\newacronym{EER}{EER}{equal error rate}

\newacronym{FHVAE}{FHVAE}{factorized hierarchical variational autoencoder}
\newacronym{FVAE}{FVAE}{factorized variational autoencoder}

\newacronym{GAP}{GAP}{global average pooling}

\newacronym{IN}{IN}{instance normalization}

\newacronym{KLD}{KLD}{Kullback-Leibler divergence}

\newacronym{MSE}{MSE}{mean squared error}

\newacronym{PER}{PER}{phone error rate}

\newacronym{SER}{SER}{speaker error rate}
\newacronym{STFT}{STFT}{short-time Fourier transform}

\newacronym{VAE}{VAE}{variational autoencoder}
\newacronym{VC}{VC}{voice conversion}
\newacronym{VTLP}{VTLP}{vocal tract length perturbation}

\def\x{{\mathbf x}}

\title{Contrastive Predictive Coding Supported Factorized Variational Autoencoder for Unsupervised Learning of Disentangled Speech Representations}
\name{Janek Ebbers\sthanks{First two authors contributed equally.
Funded by the Deutsche Forschungsgemeinschaft (DFG, German Research Foundation) - 282835863. Computational resources were provided by the Paderborn Center for Parallel Computing.}, Michael Kuhlmann$^*$, Tobias Cord-Landwehr, Reinhold Haeb-Umbach}
\address{Paderborn University, Department of Communications Engineering, Paderborn, Germany\\\texttt{\{ebbers,kuhlmann,cord,haeb\}@nt.upb.de}}
\begin{document}
\ninept
\topmargin=0mm
\setlength{\abovedisplayskip}{4pt}
\setlength{\belowdisplayskip}{4pt}
\setlength{\abovedisplayshortskip}{2pt}
\setlength{\belowdisplayshortskip}{2pt}
\setlength{\textfloatsep}{5pt}
\renewcommand\arraystretch{0.8}
\setlength{\arraycolsep}{2pt}
\maketitle
\begin{abstract}
In this work we address disentanglement of style and content in speech signals. We propose a fully convolutional variational autoencoder employing two encoders: a content encoder and a style encoder.
To foster disentanglement, we propose adversarial contrastive predictive coding.
This new disentanglement method does neither need parallel data nor any supervision.
We show that the proposed technique is capable of separating speaker and content traits into the two different representations and show competitive speaker-content disentanglement performance compared to other unsupervised approaches.
We further demonstrate an increased robustness of the content representation against a train-test mismatch compared to spectral features, when used for phone recognition.
\end{abstract}
\vspace{-1mm}

\section{Introduction}
\vspace{-1mm}
Disentangling factors of variation in data recently attracted increased interest for many modalities.
Disentangled representations are considered useful in two ways.
First, they can improve performance for various downstream tasks, which are learned on a small amount of labeled data.
In particular, it can yield improved robustness against train-test mismatches if the factors that are informative about the task can be successfully disentangled from the variations caused by a domain shift.
Second, in a disentangled representation certain factors can be modified while keeping the rest fixed, e.g., changing the lighting in an image without changing the content.
Learning disentangled representations with no or only little supervision is particularly interesting, because it can make use of the vast amounts of unannotated data available in the world.

In this paper we address disentanglement of speech signals by separating long-term variations, referred to as style, and short-term variations, referred to as content.
The proposed \gls{FVAE} employs two encoders to extract two disjoint representations, namely, a style embedding and a sequence of content embeddings, which are jointly decoded to reconstruct the input signal.\glsunset{VAE}
To restrict the content embeddings to only capture short-term variations, we propose an adversarial regularization based on \gls{CPC} \cite{oord2018representation}, which is completely unsupervised.
The basic idea is to penalize mutual information between a current and a future content embedding.
Therefore, slowly changing variations need to be captured by the style embedding.

While the proposed approach is rather general, we here focus on its capability to disentangle speaker and linguistic attributes.
Our proposed model is trained on  speech data requiring neither phonetic labels nor speaker labels.
We also do not require parallel data, where the same linguistic content is spoken by different speakers.
Compared to other unsupervised approaches, we demonstrate superior disentanglement performance in terms of downstream phone classification and speaker verification tasks.
Particularly, we show that the learned content embeddings are largely speaker invariant and achieve an increased robustness against a train-test mismatch compared to spectral features.
While we focus on disentanglement in this paper, we also provide listening examples showing that the model can be used to perform \gls{VC} when decoding content embeddings with an exchanged style embedding.

The rest of the paper is structured as follows.
After we discussed related work in Sec.~\ref{sec:related}, an overview of \gls{CPC} is given in Sec.~\ref{sec:cpc}.
Then, Sec.~\ref{sec:vae} presents our proposed \gls{FVAE} and how we use \gls{CPC} to support disentanglement.
Finally, experiments are shown in Sec.~\ref{sec:exp} and conclusions are drawn in Sec.~\ref{sec:conclusions}.

\section{Related Work}
\label{sec:related}
\vspace{-1mm}
There are many works focusing on unsupervised disentanglement of all latent factors of a generative model \cite{higgins2017beta,kim2018disentangling,chen2018isolating}.
Those works are mainly applied to toy-like image data sets, e.g., 2D shapes \cite{matthey2017dsprites}, where the generating factors are well defined.
Other works tackle supervised disentanglement of a single factor using an adversarial classifier in the latent space~\cite{lample2017fader, hadad2018two}. 

While the above works target other modalities, there are several recent works tackling disentangled speech representation learning from non-parallel data. 
Many works, e.g., \cite{hsu2016voice, van2017neural, kameoka2019acvae, chou2018multi}, focus on extracting a speaker independent content representation, while representing the speaker identity as a one-hot encoding.
Others also use speaker specific decoders \cite{polyak2019attention, lee2019voice}.
Therefore, these works can neither be used for style extraction nor for voice conversion with unknown target speakers.
Also, speaker labels are required.

Unsupervised approaches to speaker-content disentanglement are proposed in~\cite{hsu2017unsupervised, chou2019one, wu2020one, qian2019zero}.
None of these works use an explicit disentanglement objective as proposed in this paper.
The authors of~\cite{chou2019one} propose to encourage disentanglement by using \gls{IN} in the content encoder, which removes, to some extent, static signal properties such as speaker attributes.
The AutoVC model \cite{qian2019zero} relies on a carefully tuned bottleneck such that ideally all content information can be stored in the content embedding but none of the speaker-related information.
In \cite{qian2020unsupervised}, the AutoVC model was extended to an unsupervised disentanglement of timbre, pitch, rhythm and content.
The VQVC \cite{wu2020one} achieves disentanglment by applying a bottleneck in terms of vector quantization (VQ).
The \gls{FHVAE} proposed in~\cite{hsu2017unsupervised} unsupervisedly disentangles ``sequence-level'' (${>}\SI{200}{ms}$) and ``segment-level'' (${<}\SI{200}{ms}$) attributes, by restricting sequence-level embeddings to be rather static within an utterance while also putting a bottleneck on the segment-level embeddings.


\section{Contrastive Predictive Coding}
\label{sec:cpc}
Given a sequence $\mathbf{Y}=[\mathbf{y}_1,\dots, \mathbf{y}_T]$ of feature vectors $\mathbf{y}_t$, which are in our case features of a speech signal, \gls{CPC}~\cite{oord2018representation} aims at extracting the mutual information from two segments $\mathbf{Y}_t{=}[\mathbf{y}_{t-c},\dots, \mathbf{y}_{t+c}]$ and $\mathbf{Y}_{t-\tau}$, where $c$ denotes a one-sided context length and $\tau$ a shift between the segments.
For this purpose, the segments are encoded into the embeddings $\mathbf{h}_t{=}f_\text{cpc}(\mathbf{Y}_{t})$ and $\mathbf{h}_{t-\tau}{=}f_\text{cpc}(\mathbf{Y}_{t-\tau})$ such that $\mathbf{h}_{t-\tau}$ allows the prediction of the future embedding $\mathbf{h}_{t}$:
\begin{align*}
\hat{\mathbf{h}}_{t} = g_\tau(\mathbf{h}_{t-\tau})
\end{align*}
with $g_\tau(\cdot)$ denoting the projection head that predicts $\tau$ steps ahead. The \gls{CPC} model is trained using a contrastive loss \cite{oord2018representation}:
\begin{align}
\label{eq:cpc}
L_\text{cpc} = -\frac{1}{T-\tau}\sum_{t=\tau + 1}^T\frac{\exp(\mathbf{h}_t^\mathrm{T}\hat{\mathbf{h}}_t)}{\sum\limits_{\widetilde{\mathbf{h}}_t\in\mathcal{H}_t} \exp(\widetilde{\mathbf{h}}_t^\mathrm{T}\hat{\mathbf{h}}_t)}\,\,,
\end{align}
where $\mathcal{H}_t$ is the set of candidate embeddings ${\{\mathbf{h}^{(b)}_t\,|\, 1\leq b\leq B\}}$ in the mini-batch of size $B$.
Note that Eq. \eqref{eq:cpc} equals a cross entropy loss including a softmax where the logits are given as the inner product of the predicted embedding $\hat{\mathbf{h}}_t$ and the candidate embeddings $\widetilde{\mathbf{h}}_t\in \mathcal{H}_t$.
Hence, the model is essentially trained to be able to correctly classify the true future embedding out of a couple of candidates.
The number of steps $\tau$ that the model predicts into the future controls the kind of mutual information that is encoded.
If $\tau$ is small, i.e., the segments are very close to each other, the model probably learns to recognize content attributes, e.g., whether they are parts of the same acoustic unit.
However, if $\tau$ is large, i.e., the segments are further apart, the mutual information the model has to recognize are primarily the static properties such as speaker attributes.

\section{Contrastive Predictive Coding Supported Factorized Variational Autoencoder}
\label{sec:vae}
\begin{figure}[t]
	\centering
	\resizebox{0.45\textwidth}{!}{
		\begin{tikzpicture}[>=stealth',semithick,auto,
		scale=1, 
		block/.style={
			rectangle,
			draw,
			text centered,
		},
		encoder/.style={
			trapezium,
			draw,
			trapezium angle=70,
			rotate=-90,
			text centered,
			minimum height=1cm,
			minimum width=1cm,
			trapezium stretches=true
		},
		decoder/.style={
			trapezium,
			draw,
			trapezium angle=70,
			rotate=90,
			text centered,
			minimum height=1cm,
			minimum width=1cm,
			text width=.6cm,
			trapezium stretches=true,
			anchor=bottom left corner
		},
		arrow/.style={
			thick,
			->,
			>=stealth
		}
		]

		\newcommand{\encoderSpacing}{2.2}
		\newcommand{\decoderDistance}{2.0}
		\newcommand{\decoderOffset}{0.9}
		\newcommand{\sampleOffset}{.0}
		\newcommand{\InputSpacing}{.4}

		\node [encoder] (sEncoder) {\rotatebox[]{90}{\stackanchor{\phantom{y}\phantom{y}Style\phantom{y}\phantom{y}}{Encoder}}};
		\coordinate (sEncoder tlc) at (sEncoder.bottom left corner);
		\coordinate (sEncoder blc) at (sEncoder.bottom right corner);
		\coordinate (sEncoder trc) at (sEncoder.top left corner);
		\coordinate (sEncoder brc) at (sEncoder.top right corner);

		\node [encoder, below=\encoderSpacing of sEncoder.bottom left corner, anchor=bottom left corner] (ZEncoder) {\rotatebox[]{90}{\stackanchor{\phantom{y}Content\phantom{y}}{Encoder}}};
		\coordinate (ZEncoder tlc) at (ZEncoder.bottom left corner);
		\coordinate (ZEncoder blc) at (ZEncoder.bottom right corner);
		\coordinate (ZEncoder trc) at (ZEncoder.top left corner);
		\coordinate (ZEncoder brc) at (ZEncoder.top right corner);

		\path (sEncoder blc) -- (sEncoder tlc) node[midway, left=.8cm] (X2) {$\mathbf{X}$};

		\path (sEncoder tlc) -- (sEncoder blc) node[midway, shape=coordinate] (sEnc-in) {};
		\path (ZEncoder tlc) -- (ZEncoder blc) node[midway, shape=coordinate] (ZEnc-in) {};
		\path (sEncoder trc) -- (sEncoder brc) node[midway, shape=coordinate] (sEnc-out) {};
		\path (ZEncoder trc) -- (ZEncoder brc) node[midway, shape=coordinate] (ZEnc-out) {};
		\path (sEnc-out) -- (ZEnc-out) node[midway, shape=coordinate] (Enc-out-center) {};

		\node [decoder, anchor=north] at ($(Enc-out-center) + (\decoderDistance, \decoderOffset)$) (Decoder) {\rotatebox[]{-90}{Decoder}};

		\coordinate (Decoder trc) at (Decoder.bottom right corner);
		\coordinate (Decoder brc) at (Decoder.bottom left corner);
		\coordinate (Decoder tlc) at (Decoder.top right corner);
		\coordinate (Decoder blc) at (Decoder.top left corner);

		\path (Decoder tlc) -- (Decoder blc) node[midway, shape=coordinate] (Dec-in) {};
		\path (Decoder trc) -- (Decoder brc) node[midway, left] (Dec-out) {};
		\node [right=.6cm of Dec-out] (reconstruction) {\( \hat{\mathbf{X}} \)};

		\path let
		\p{A} = (X2),
		\p{B} = (ZEncoder)
		in
		coordinate (norm) at ( \x{A}, \y{B});
		\node[block, anchor=center] at (norm) (norm) {\stackanchor{Instance}{Norm}};

		\path (X2) -- (norm) node[midway, shape=coordinate] (vtlp) {};
		\node[block] at (vtlp) (vtlp) {VTLP};

		\node[block, anchor=center] at ($(sEnc-out) + (\decoderDistance/2+\sampleOffset, 0.)$) (gap) {GAP};
		\node[block, anchor=center] at ($(ZEnc-out) + (\decoderDistance/2+\sampleOffset, \encoderSpacing/4-\InputSpacing/4+\decoderOffset/2 + .05)$) (sample) {Sample};
		\path [{}-{>}, draw]
		(X2) edge[] node[]{} (sEnc-in)
		(X2) edge[] node[]{} (vtlp)
		(vtlp) edge[] node[]{} (norm)
		(norm) edge[] node[]{} (ZEnc-in)
		(Dec-out) edge[] node[]{} (reconstruction)
		(sEnc-out) edge[] node[pos=0.45]{\(\mathbf{S}\)} (gap)
		(gap) -- node[pos=0.45]{\(\bar{\mathbf{s}}\)} ($(Dec-in) + (0, \InputSpacing/2)$);
		\path [{}-{>}, draw]
		(ZEnc-out) -- ++ (\decoderDistance/2, 0) node[pos=0.5, below]{\( q(\mathbf{Z}) \)} -- (sample);
		\path [{}-{>}, draw]
		(sample) -- ($(Dec-in) + (-\decoderDistance/2+\sampleOffset, -\InputSpacing/2)$) -- node[pos=.7, below]{\( \mathbf{Z} \)} ($(Dec-in) + (0, -\InputSpacing/2)$);
		\end{tikzpicture}
	}
	\caption{Factorized VAE.}
	\vspace{-1mm}
	\label{fig:fvae}
\end{figure}
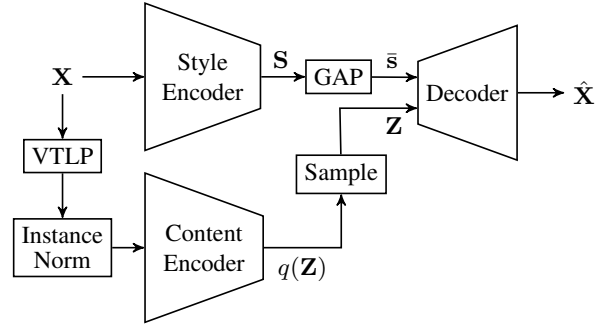

To learn disentangled representations of style and content, we propose a fully convolutional \gls{FVAE} which is illustrated in Fig. \ref{fig:fvae}.
The \gls{FVAE} employs two encoders: a content encoder to encode content information from an input sequence~$\mathbf{X}{=}[\mathbf{x}_1,\dots,\mathbf{x}_T]$ into a sequence of content embeddings~$\mathbf{Z}{=}[\mathbf{z}_1,\dots,\mathbf{z}_N]$, and a style encoder to extract style traits into a style embedding~$\bar{\mathbf{s}}$.

\FloatBarrier

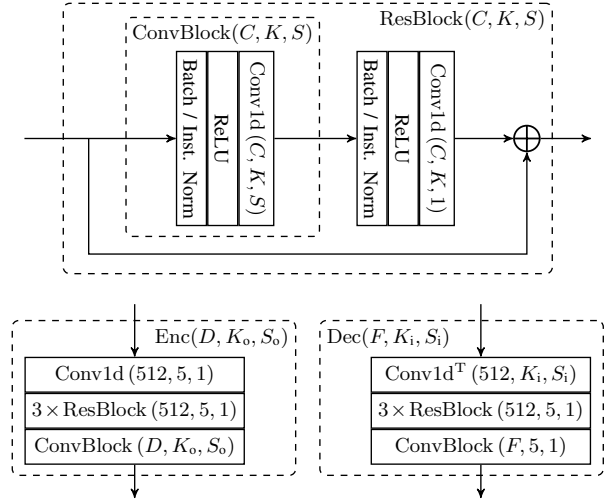
\begin{figure}[t]
	\centering
	\resizebox{0.45\textwidth}{!}{
		\begin{tikzpicture}[>=stealth',semithick,auto,
		scale=1,
		vblock/.style={
			rectangle,
			draw,
			text centered,
			rotate=-90,
			text width=2.5cm,
			text height=0.25cm
		},
		hblock/.style={
			rectangle,
			draw,
			text centered,
			text width=3.2cm,
			text height=0.25cm
		},
		arrow/.style={
			thick,
			->,
			>=stealth
		},
		sum/.style={
		path picture={\draw[black](path picture bounding box.south) -- (path picture bounding box.north) (path picture bounding box.west) -- (path picture bounding box.east);}
		}
		]

		\node [vblock] at (2.6,0) (norm1) {Batch / Inst. Norm};
		\node [vblock, above=0 of norm1] (relu1) {ReLU};
		\node [vblock, above=0 of relu1] (conv1) {$\mathrm{Conv1d}\,(C,K,S)$};

		\node [vblock] at ($(conv1)+(1.8,0)$) (norm2) {Batch / Inst. Norm};
		\node [vblock, above=0 of norm2] (relu2) {ReLU};
		\node [vblock, above=0 of relu2] (conv2) {$\mathrm{Conv1d}\,(C,K,1)$};

		\node[draw, circle, sum, thick, minimum width = 0.4cm] at ($(conv2) + (1.4, 0)$) (sum){};

		\path [{}-{>}, draw]
		(0,0) -- (norm1);
		\path [{}-{>}, draw]
		(conv1) edge[] node[]{} (norm2)
		(conv2) edge[] node[]{} (sum)
		(sum) edge[] node[]{} ($(sum) + (1, 0)$);
		\path [{}-{>}, draw]
		(1,0) -- (1,-1.8) -- ($(sum) + (0, -1.8)$) -- (sum);

		\node[draw, dashed, rounded corners=0.1cm, text width=7.4cm, text height=4cm] at ($(conv1) + (0.8, 0)$) (box) {};
		\node[anchor=north east] at (box.north east) () {$\mathrm{ResBlock}(C,K,S)$};
		\node[draw, dashed, rounded corners=0.1cm, text width=2.8cm, text height=3.2cm] at ($(relu1)+(0,0.18)$) (box2) {};
		\node[anchor=north] at (box2.north) () {$\mathrm{ConvBlock}(C,K,S)$};

		\node [hblock, anchor=north west] at (0,-3.4) (EncConvIn) {$\mathrm{Conv1d}\,(512,5,1)$};
		\node [hblock, below=0 of EncConvIn] (EncResBlocks) {$3\times\mathrm{ResBlock}\,(512,5,1)$};
		\node [hblock, below=0 of EncResBlocks] (EncConvOut) {$\mathrm{ConvBlock}\,(D,K_\text{o},S_\text{o})$};
		\path [{}-{>}, draw] ($(EncConvIn) + (0, 1.1)$) -- (EncConvIn);
		\path [{}-{>}, draw] (EncConvOut) -- ($(EncConvOut) - (0, .8)$);
		\node[draw, dashed, rounded corners=0.1cm, text width=4.2cm, text height=2.2cm] at ($(EncResBlocks) + (0.3, 0.2)$) (boxEnc) {};
		\node[anchor=north east] at (boxEnc.north east) () {$\mathrm{Enc}(D,K_\text{o},S_\text{o})$};

		\node [hblock, anchor=north east] at ($(sum) + (1, -3.4)$) (DecConvIn) {$\mathrm{Conv1d^\mathrm{T}}\,(512,K_\text{i},S_\text{i})$};
		\node [hblock, below=0 of DecConvIn] (DecResBlocks) {$3\times\mathrm{ResBlock}\,(512,5,1)$};
		\node [hblock, below=0 of DecResBlocks] (DecConvOut) {$\mathrm{ConvBlock}\,(F,5,1)$};

		\path [{}-{>}, draw] ($(DecConvIn) + (0, 1.1)$) -- (DecConvIn);
		\path [{}-{>}, draw] (DecConvOut) -- ($(DecConvOut) - (0, .8)$);
		\node[draw, dashed, rounded corners=0.1cm, text width=4.2cm, text height=2.2cm] at ($(DecResBlocks) + (-0.3, 0.2)$) (boxDec) {};
		\node[anchor=north west] at (boxDec.north west) () {$\mathrm{Dec}(F,K_\text{i},S_\text{i})$};

		\end{tikzpicture}
	}
	\caption{CNN Architectures: $\mathrm{Conv1d}\,(C,K,S)$ denotes a one-dimensional convolutional layer with $C$ output channels, kernel size of $K$ and striding of $S$. $\mathrm{Conv1d^\mathrm{T}}$ denotes a  one-dimensional transposed convolutional layer.}
	\label{fig:cnn}
\end{figure}

As input signal representation $\mathbf{X}$ we extract $F{=}80$ log mel-filter bank energy (F-Bank) features for each frame of a \gls{STFT} using an audio sample rate of \SI{16}{kHz}, a frame length of \SI{50}{ms} and a hop-size of \SI{12,5}{ms}.
Each log-mel band is normalized by subtracting the global mean and dividing by the global standard deviation, which are determined on a training set.

The encoders and the decoder are one-dimensional \glspl{CNN} as shown in Fig. \ref{fig:cnn}.
The style encoder uses $K_\text{o}{=}S_\text{o}{=}1$ yielding frame level outputs~$\mathbf{S}{=}[\mathbf{s}_1,\dots,\mathbf{s}_T]$ on which a \gls{GAP} over time is applied to obtain an utterance-level style embedding $\bar{\mathbf{s}}$.
The content encoder uses $K_\text{o}{=}S_\text{o}{=}S_\text{ds}$ with downsampling being performed when $S_\text{ds}{>}1$ such that $N{=}\lceil T/S_\text{ds}\rceil$.
The concatenated embeddings 
\( \begin{bmatrix} 
	\bar{\mathbf{s}} & \dots & \bar{\mathbf{s}} \\
	\mathbf{z}_1 & \dots & \mathbf{z}_{N}
\end{bmatrix} \) 
are forwarded to the decoder network which outputs an input signal reconstruction $\hat{\mathbf{X}}$.
The input layer of the decoder upsamples the embeddings of length $N$ back to length $T$ by using $K_\text{i}{=}S_\text{i}{=}S_\text{ds}$.
The training objective for reconstruction is given by the \gls{MSE}:
\begin{align*}
L_\text{rec} = \frac{1}{T}||\hat{\mathbf{X}} - \mathbf{X}||_2^2\,\,.
\end{align*}

To support style extraction, we apply an auxiliary \gls{CPC} loss $L^{(\mathbf{S})}_\text{cpc}$ to the style encoder output, which is the loss from Eq.~\eqref{eq:cpc} with variables $\mathbf{h}_t$ replaced by $\mathbf{s}_t$ here.
As we aim to extract utterance-level style information, we choose a large $\tau{=}80$ which corresponds to a shift of \SI{1}{s}.
Furthermore, the projection head $g_\tau(\cdot)$ is chosen to be the identity $\hat{\mathbf{s}}_{t}{=}\mathbf{s}_{t-\tau}$ such that the style encoder is encouraged to extract similar embeddings within the same utterance and orthogonal embeddings in different utterances.

Naturally, the proposed model would tend to access all the required signal information through $\mathbf{Z}$ while ignoring $\mathbf{s}$ as there is usually much more capacity in a sequence of embeddings~$\mathbf{Z}$ than in a single embedding~$\bar{\mathbf{s}}$.
Therefore, the challenge is to prevent the model from also encoding style properties of the signal into $\mathbf{Z}$ but make the model access it through $\bar{\mathbf{s}}$.

\Glspl{VAE} \cite{kingma2014auto} have been used to put an information bottleneck on the content embedding which has shown to improve disentanglement \cite{hsu2016voice}.
Here, $\mathbf{z}_n$ is interpreted as a stochastic variable with prior $p(\mathbf{z}_n){=}\mathcal{N}(\mathbf{z}_n; \mathbf{0}, \mathbf{I})$ and an approximate posterior $q(\mathbf{z}_n){=}\mathcal{N}(\mathbf{z}_n; \boldsymbol{\mu}_n, \mathrm{diag}(\boldsymbol{\sigma}_n^2))$, with the content encoder providing $\boldsymbol{\mu}_n$ and $\log\left(\boldsymbol{\sigma}_n^2\right)$.
The content embeddings that are forwarded into the decoder are sampled as $\mathbf{z}_n{\sim}q(\mathbf{z}_n)$ using the reparameterization trick \cite{kingma2014auto} during training, while being set to $\mathbf{z}_n{=}\boldsymbol{\mu}_n$ in test mode.
The \gls{KLD}
\begin{align*}
L_\text{kld} &= \frac{1}{N} \sum_{n=1}^N \mathrm{KL}(q(\mathbf{z}_n)||p(\mathbf{z}_n))
\end{align*}
that is added to the \gls{VAE} objective prefers the posterior $q(\mathbf{z}_n)$ to be uninformative which helps encoding information into $\bar{\mathbf{s}}$ rather than~$\mathbf{Z}$.
However, it also harms reconstruction which is why it should not be weighted too high either.

Another simple additional measure to encourage disentanglement is to remove or distort utterance-level properties in the input of the content encoder.
Then, $\mathbf{X}$ cannot be reconstructed without accessing information via the style encoder.
First, during training we distort speaker properties using \gls{VTLP} \cite{jaitly2013vocal}.
\gls{VTLP} was proposed to increase speaker variability when training speech recognition systems.
For this purpose, the center bins of the mel-filter-banks are randomly remapped using a piece-wise linear warping function.
Second, we perform \gls{IN} \cite{ulyanov2016instance} on the content input, i.e., the individual log-mel bands of each input signal are normalized across time to zero mean and unit variance.
We perform \gls{IN} also in the hidden layers of the content encoder, instead of batch normalization used in the other sub-networks, which has been found useful to encourage speaker-content disentanglement~\cite{chou2019one}.

However, none of the above measures explicitly prevents the content encoder from encoding style.
To enforce disentanglement, the authors of \cite{chou2018multi} suggested to employ a jointly trained adversarial speaker classifier on content embeddings to make the content encoder drop information revealing the speaker identity.
However, the adversarial speaker classifier has two main disadvantages.
First, although it does not require text annotations, it still requires speaker annotations.
Second, it does not scale to large unbalanced databases with a huge number of speakers as the classification task itself becomes too uncertain to obtain useful adversarial gradients.

Therefore, in this work we propose adversarial \gls{CPC} as an alternative which is fully unsupervised and independent of the (unobserved) number of speakers.
Hence, this approach has the potential to scale to large unlabeled databases.

We train a \gls{CPC} encoder $\mathbf{H}{=}[\mathbf{h}_1,\dots,\mathbf{h}_T]{=}f_\text{acpc}\left(\mathbf{Y}_\mathbf{Z}\right)$ on the content encoder's output \(\mathbf{Y}_\mathbf{Z}{=}[\mathbf{y}_{\mathbf{z}_1},\dots,\mathbf{y}_{\mathbf{z}_N}]{=}\begin{bmatrix} 
	\mathbf{M}_\mathbf{Z} \\
	\log\mathbf{V}_\mathbf{Z}
\end{bmatrix}\)
with $\mathbf{M}_\mathbf{Z}{=}[\boldsymbol{\mu}_{\mathbf{z}_1},\dots, \boldsymbol{\mu}_{\mathbf{z}_N}]$ being the sequence of posterior means and $\mathbf{V}_\mathbf{Z}{=}[\boldsymbol{\sigma}^2_{\mathbf{z}_1},\dots, \boldsymbol{\sigma}^2_{\mathbf{z}_N}]$ being the sequence of posterior variances.
For the \gls{CPC} encoder we use the architecture $f_\text{cpc}{=}\mathrm{Dec}(D_\mathbf{h}, K_\text{ds}, S_\text{ds})$ similar to the \gls{VAE} decoder with $D_\mathbf{h}{=}128$.
As before, we are interested in extracting global style information which is why we again choose $\tau{=}80$ corresponding to \SI{1}{s} and an identity projection $\hat{\mathbf{h}}_{t}{=}\mathbf{h}_{t-\tau}$.
The \gls{CPC} encoder is trained to minimize the \gls{CPC} loss~\eqref{eq:cpc}, which is referred to as $L^{(\mathbf{Z})}_\text{cpc}$ here.
By adding the negative of the \gls{CPC} loss to the \gls{FVAE} objective, the content encoder tries to maximize it, i.e., the \gls{FVAE} content encoder and the \gls{CPC} encoder operate adversarially here.
This prevents the \gls{FVAE} content encoder from encoding mutual information into the content embeddings which are \SI{1}{s} apart (or further) eventually preventing it from encoding speaker attributes and other static style properties.

Finally, our complete training objective is given as
\begin{align*}
L_\text{fvae} &= L_\text{rec} + \beta L_\text{kld} + \lambda_\mathbf{S}L^{(\mathbf{S})}_\text{cpc} - \lambda_\mathbf{Z}L^{(\mathbf{Z})}_\text{cpc}\,\,.
\end{align*}

\section{Experiments}
\label{sec:exp}
Experiments are performed on the Librispeech corpus~\cite{panayotov2015librispeech} which is derived from audiobooks and contains, for each speaker, utterances from one or multiple chapters.
Here, \textit{train-clean-100} and \textit{train-clean-360} subsets are used for training.
These sets contain \SI{100}{h} and \SI{360}{h} of clean speech from 251 and 921 speakers, respectively.
From \textit{train-clean-100}, however, we only use \SI{60}{\%} of the speakers' utterances for training and leave $2{\times}\SI{20}{\%}$ for validation and evaluation of a closed-speaker scenario.
In the following we therefore refer to the train, validation and test portions as \textit{train-clean-60}, \textit{dev-closed} and \textit{test-closed}, respectively.
Validation and evaluation of an open-speaker scenario is done with the official \textit{dev-clean} and \textit{test-clean} sets, respectively.
In all trainings, utterances shorter than \SI{2}{s} are discarded.
Remaining utterances are cut into equally long segments such that no segment is longer than \SI{4}{s}.

We compare our proposed method with the following disentanglement baselines that are also scalable and can be trained fully unsupervised, i.e., also without any supervision on speaker labels.

\textit{\Gls{AdaIN}:}
The authors of \cite{chou2019one} proposed to use \gls{IN} in the content encoder to normalize style information. 
The decoder uses AdaIN \cite{huang2017arbitrary} which applies \gls{IN} and then adaptively computes shifts and scales of an affine transformation from a style embedding extracted by a style encoder.

\textit{Dimensionality bottleneck:}
We already motivated that it is crucial to put an information bottleneck on the content embeddings to achieve disentanglement.
The authors of \cite{qian2019zero} found that disentanglement can be achieved by only carefully tuning the dimensionality of the latent content representation.
We therefore compare our proposed \gls{CPC} supported \gls{FVAE} with both the AutoVC model proposed in \cite{qian2019zero} as well as with a simple dimensionality tuning in our proposed \gls{FVAE} without applying any supporting \gls{CPC} losses.
AutoVC usually uses a pre-trained speaker encoder trained using speaker labels.
To make the model fully unsupervised for a fair comparison, we here replace it  with a vanilla CPC encoder $\mathbf{H}=f_\text{cpc}(\mathbf{X})$ pre-trained on F-bank features $\mathbf{X}$ with $f_\text{cpc}=\mathrm{Enc}(128, 1, 1)$, $\tau{=}80$ and $g_\tau(\mathbf{h})=\mathbf{h}$ to extract style embeddings.

\gls{FVAE} and baseline models are trained on the combined \textit{train-clean-60} and \textit{train-clean-360} sets for $10^5$ update steps using mini-batches of 32 segments.
When using an adversarial CPC model, it is exclusively updated for three additional steps after each joint update of \gls{FVAE} and \gls{CPC}.
Further, an initial warm up of 400 exclusive \gls{FVAE} steps followed by 1200 exclusive adversarial \gls{CPC} steps is performed.
Adam \cite{kingma2014adam} is used for optimization with a learning rate of $5{\cdot}10^{-4}$ and gradient clipping is applied using thresholds of 10, 20 and 2 for VAE encoders, decoder and CPC encoder, respectively.
After training, the checkpoint which achieves lowest reconstruction error on the validation set is used to report results on the test set.
For the AdaIN and AutoVC baselines we use the model implementations released by the authors.
Content and speaker embedding sizes of $D_\mathbf{z}{=}32$ and $D_\mathbf{s}{=}128$ are used for all models.

The vanilla \gls{CPC} model is, similarly to the disentanglement models, trained on the combined \textit{train-clean-60} and \textit{train-clean-360} sets for $5\cdot 10^4$ update steps using mini-batches of 64, a learning rate of $10^{-3}$ and gradient clipping at a threshold of 2.

Disentanglement performance is evaluated as follows.

\textit{Post-hoc classifiers:}
After training, downstream phone and speaker classifiers are trained on the models' content embeddings $\mathbf{Z}$.
As the content embeddings are supposed to be informative about the linguistic content, we aim to obtain a low \gls{PER}.
In contrast, they are also expected to be speaker invariant which should therefore not allow speaker classification.
Hence, a high \gls{SER} indicates good disentanglement.
Similar evaluations were made, e.g., in \cite{chou2019one,qian2019zero}.
The architecture of the classifiers is similar to the decoder with $f^{(\mathbf{Z})}_\text{phn}{=}\mathrm{Dec}(40,S_\text{ds},S_\text{ds})$ and $f^{(\mathbf{Z})}_\text{spk}{=}\mathrm{Dec}(251,S_\text{ds},S_\text{ds})$, where $40$ is the number of target phones and $251$ the number of target speakers.
If not stated otherwise, the classifiers are trained on the \textit{train-clean-60} set with the phone classifier being validated and evaluated on \textit{dev-clean} and \textit{test-clean} and the speaker classifier on \textit{dev-closed} and \textit{test-closed}, respectively.
Phone alignments for training are imported from Kaldi \cite{povey2011kaldi}.
Training is performed for $5\cdot 10^4$ update steps using mini-batches of 64, a learning rate of $10^{-3}$ and gradient clipping at a threshold of 20.
The checkpoint which achieves highest accuracy on the validation set is used to report results on the test set.
As reference, we also report performance of classifiers trained with F-bank features.

\textit{Speaker Verification:}
Speaker verification aims to recognize a pair of utterances as either from the same speaker or from different speakers by evaluating the similarity of style embeddings.
Here, the distance between two embeddings is measured by their cosine distance where a pair is classified as same speaker when the distance falls below a certain threshold.
Performance is reported as the \gls{EER} which is the error rate for the threshold where the false acceptance rate and false rejection rate are equal.
We aim at a low \gls{EER} as we expect to have similar style embeddings for utterances from the same speaker and different ones for utterances from different speakers.
As Librispeech is not a standard set for speaker verification, we randomly generate a verification task.
For each utterance from \textit{test-clean} we sample an utterance from every other speaker to generate a negative pair yielding a total of ${\sim}\SI{100}{k}$ negative trials.
For positive pairs we consider two setups:
In the \textit{within chapter (WC)} setup all possible combinations of a speakers' utterances within the same chapter are considered totaling ${\sim}\SI{50}{k}$ positive pairs.
In the \textit{across chapter (AC)} setup we randomly replace one of the utterances in each WC pair, if possible\footnote{Some speakers only have utterances from a single chapter.}, by an utterance from another chapter.
As reference we report the performance of i-vectors~\cite{dehak2010front} which have been trained on \textit{train-clean-60} using Kaldi.
Note that i-vector training requires speaker supervision though and can hence be considered a topline.
We further report performance of the vanilla \gls{CPC} model which is unsupervised.

\begin{table}[t]
	\caption{Disentanglement performance measured by downstream PER on the content embedding (lower is better), SER on the content embedding (higher is better) and speaker verification on the style embedding (lower is better). All values are stated in \%. Best result in a block is shown in bold.}
	\vspace{-.5mm}
	\label{tab:results-hyperparams}
	\centering
	\resizebox{0.46\textwidth}{!}{
	\setlength{\tabcolsep}{2pt}
	\begin{tabular}{cc|ccccc|cccc}
		\toprule[1.5pt]
		\multirow{2}{*}{\#} & \multirow{2}{*}{Model} & \multirow{3}{*}{IN} & \multirow{3}{*}{VTLP} & \multirow{3}{*}{$S_\text{ds}$} & \multirow{3}{*}{$\lambda_\mathbf{Z}$} & \multirow{3}{*}{$\lambda_\mathbf{S}$} & \multirow{2}{*}{PER($\mathbf{Z}$)} & \multirow{2}{*}{SER($\mathbf{Z}$)} & \multicolumn{2}{c}{\,EER($\bar{\mathbf{s}}$)}\\
		& & & & & & & & & WC & AC\\
		\midrule
		1& {F-Bank} & \xmark & \xmark & - & - & - & 17.7 & 1.7 & 17.3 & 22.4 \\
		2& {F-Bank} & \cmark & \cmark & - & - & - & 15.3 & - & - & - \\
		3& {I-Vector} & - & - & - & - & - & - & - & 4.1 & 6.7 \\
		4& {CPC} & \xmark & \xmark & - & - & - & - & - & 2.4 & 6.7 \\
		\midrule
		5& {AdaIN \cite{chou2019one}} & \cmark & \xmark & 8 & - & - & \textbf{21.4} & 37.4 & 7.8 & 11.2 \\
		6& {AutoVC \cite{qian2019zero}} & \xmark & \xmark & 16 & - & - & 22.0 & \textbf{52.9} & \textbf{2.4}$^*$ & \textbf{6.7}$^*$ \\
		\midrule
		7& FVAE & \cmark & \cmark & 8 & 0 & 0 & \textbf{16.6} & 18.8 & 10.1 & 13.9 \\
		8& FVAE & \cmark & \cmark & 16 & 0 & 0 & 17.8 & 38.0 & 9.2 & 12.9 \\
		9& FVAE & \cmark & \cmark & 32 & 0 & 0 & 19.8 & 65.6 & 8.8 & 12.2 \\
		10& FVAE & \cmark & \cmark & 64 & 0 & 0 & 28.3 & \textbf{86.2} & \textbf{6.6} & \textbf{10.0} \\
		\midrule
		11& FVAE & \cmark & \cmark & 8 & 1 & 0 & 17.4 & \textbf{51.3} & 7.9 & 11.8 \\
		12& FVAE & \cmark & \cmark & 8 & 0 & 1 & \textbf{16.7} & 17.3 & 2.2 & \textbf{6.6} \\
		13& FVAE & \cmark & \cmark & 8 & 1 & 1 & 17.4 & 48.1 & 2.1 & 6.7 \\
		14& FVAE & \cmark & \xmark & 8 & 1 & 1 & 17.8 & 44.0 & 2.2 & 6.7 \\
		15& FVAE & \xmark & \xmark & 8 & 1 & 1 & 20.0 & 28.6 & \textbf{2.0} & 6.7 \\
		\bottomrule[1.5pt]
	\end{tabular}
	}
	\phantom{{\tiny}}\\
	\vspace{.5mm}\hspace{-1cm}{\footnotesize \qquad\qquad $^*$Scores from pre-trained \gls{CPC} which is used by AutoVC as style encoder.}
\end{table}

Table~\ref{tab:results-hyperparams} shows results for different models and different values for the hyper parameters $S_\text{ds}$ (subsampling factor determining the dimensionality bottlenck), $\lambda_\mathbf{Z}$ (adversarial \gls{CPC} weight), $\lambda_\mathbf{S}$ (auxiliary \gls{CPC} weight), usage of \gls{IN} and usage of \gls{VTLP}.
All \glspl{FVAE} have been trained using $\beta{=}0.01$.
The first and second block show reference and baseline systems, respectively.
The third block presents a dimensionality tuning in our proposed architecture without \gls{CPC} support which can also be viewed a baseline for the proposed additional \gls{CPC} losses.
The last block reports results when using \gls{CPC} supported training.


Evaluation of the content embeddings shows that all models are able to significantly improve speaker invariance compared to F-bank features.
Model 10 gives the best \gls{SER} result achieving 86.2\%.
However, this comes with the cost of a high \gls{PER}.
Comparing \glspl{PER}, it can be seen that in most configurations the proposed \gls{FVAE} significantly outperforms the baselines.
In block three it can be seen that a smaller bottleneck due to a higher subsampling factor leads to an increase in both \gls{PER} and \gls{SER} as expected.
Here, model 8 gives a good balance between \gls{PER}, which is still not far off, and speaker invariance, which is decently high.
Compared to model 8, the models 11 and 13 from block four use a smaller $S_\text{ds}$ but use adversarial \gls{CPC}.
It can be seen that this yields an improvement of both \gls{PER} and speaker invariance.
Comparing the last three models 13, 14 and 15 shows that \gls{PER} and speaker invariance both highly benefit from \gls{IN} and slightly benefit from \gls{VTLP}.

Moving on to the evaluation of the style embeddings, it can be seen that WC verification generally achieves a significantly higher performance than AC verification.
This can be explained due to differences in recording conditions between utterances from different chapters, such as volume and reverberation.
Note that unsupervisedly trained style encoders are meant to capture such style differences and, hence, a worse verification performance in the AC setup is not necessarily a sign of bad quality here.
It can be seen that in both setups the best verification performance can be achieved by our proposed \gls{FVAE} when using an auxiliary \gls{CPC} loss on the style encoders' output (models 12-15) allowing to even outperform i-vectors and vanilla CPC style embeddings.

Next, we evaluate the robustness of the proposed content feature extraction in the case of little training data suffering from a train-test mismatch.
Such a scenario is simulated by training a phone classifier on only a subset of $20$ male speakers from \textit{train-clean-60} and test on female speakers.
Results are reported in Tab.~\ref{tab:results-mismatched} for both, when using VTLP and when not using VTLP in classifier training as well as in \gls{FVAE} training (models 13,14 in Tab.\ref{tab:results-hyperparams}).
It can be seen that our proposed model outperforms F-Bank features in mismatched evaluations.
As expected the performance difference is, however, much smaller when VTLP is used.
However, note that the male-female mismatch is only a proof of concept.
Other mismatches might not be compensated by a simple data augmentation such as VTLP.

\begin{table}[t]
	\caption{Mismatched \gls{PER} when only trained on male speakers. Best result in column is shown bold.}
	\vspace{-.5mm}
	\label{tab:results-mismatched}
	\centering
	\setlength{\tabcolsep}{2pt}
	\resizebox{0.32\textwidth}{!}{
	\begin{tabular}{c|ccc|ccc}
		\toprule[1.5pt]
		\multirow{2}{*}{Features} & \multicolumn{3}{c|}{w/o VTLP} & \multicolumn{3}{c}{w/ VTLP}\\
		& male & female & all & male & female & all \\
		\midrule
		{F-Bank} & \textbf{24.5} & 42.8 & 33.5 & \textbf{23.4} & 30.2 & 26.7 \\
		{FVAE} & 24.8 & \textbf{31.14} & \textbf{28.2} & 24.5 & \textbf{28.4} & \textbf{26.4}\\
		\bottomrule[1.5pt]
	\end{tabular}
	}
\end{table}

Finally, we provide listening examples\footnote{\url{go.upb.de/acpcvc}} showing that the model is able to perform zero-shot \gls{VC}.
An extensive investigation and evaluation of the model's \gls{VC} capabilities is, however, left to the future.

\section{Conclusions}
\label{sec:conclusions}
The proposed \gls{CPC} support for training an \gls{FVAE} conducts disentanglement of speaker and content-induced variations.
Its training is fully unsupervised and does not even require knowledge of speaker labels.
Compared to other unsupervised disentanglement approaches, superior disentanglement can be achieved in terms of down stream phone recognition and speaker verification.
Further, the proposed content embedding extraction allows to obtain an increased robustness against a train-test mismatch when used for phone classification.

\balance
\bibliographystyle{IEEEbib}
\bibliography{strings,refs}

\end{document}